\documentclass[preprint,12pt]{elsarticle}
\makeatletter
\def\ps@pprintTitle{%
 \let\@oddhead\@empty
 \let\@evenhead\@empty
 \def\@oddfoot{}%
 \let\@evenfoot\@oddfoot}
\makeatother

\usepackage{amssymb}

\usepackage{amsmath}%
\newdefinition{definition}{Definition}

\newproof{pf}{Proof}
\usepackage{booktabs,tabu}
\usepackage{caption}
\usepackage{subcaption}
\usepackage{hyperref}
\hypersetup{ 
colorlinks,
linkcolor=blue,
urlcolor=blue,
}
\usepackage{algorithm}
\usepackage[noend]{algpseudocode}

\begin{document}

\begin{frontmatter}



\title{A Note on Particle Gibbs Method \\and \\its Extensions and Variants}



\author[label1]{Niharika Gauraha}
 \address{Division of Computational Science and Technology \\KTH Royal Institute of Technology, Sweden }
 \ead{niharika@kth.se}

\begin{abstract}
High-dimensional state trajectories of state-space models pose challenges for Bayesian inference. 
Particle Gibbs (PG) methods have been widely used to sample from the posterior of a state space model. Basically, particle Gibbs is a Particle Markov Chain Monte Carlo (PMCMC) algorithm that mimics the Gibbs sampler by drawing model parameters and states from their conditional distributions.

This tutorial provides an introductory view on Particle Gibbs (PG) method and its extensions and variants, and illustrates through several examples of inference in non-linear state space models (SSMs). We also implement PG Samplers in two different programming languages: Python and Rust. Comparison of run-time performance of  Python and Rust programs are also provided for various PG methods.

\end{abstract}

\end{frontmatter}


\section{Introduction} \label{sec:intro}
State-space models (SSMs)  have been used extensively to model time series and dynamical systems. 
The SSMs can be broadly divided into two groups, linear Gaussian and nonlinear and/or non-Gaussian. In this tutorial, we mainly focus on the later group nonlinear SSMs as defined below.
\begin{subequations}
\begin{align}
x_t &= f (x_{t-1} ) + \epsilon_t, \quad x_t \in R^{n_x} \\
y_t &= g(x_t ) + w_t, \quad y_t \in R^{n_y} \\
x_1 & \sim p(x_1 ),
\end{align}
\end{subequations} \label{eq:ssm}
where the system noise $\epsilon_t \sim N (0, Q)$ and the measurement noise $w_t \sim N (0, R)$ are both Gaussian. The variables $x_t$ for $t = 1, \ldots, T$ are latent (unobserved) variables and  $y_t$ for $t = 1, \ldots, T$ are observed variables. The functional form of $f$ and $g$ are assumed to be known. Usually,  learning of a SSM involves the parameter inference problem as well as the state inference problem. More specifically, we are concerned with the probabilistic learning of SSMs, by inferring noise variances $Q$ and $R$, along with the states trajectories $x_t$ for $t = 1, \ldots, T$ conditioned on the given $T$ observations $y_{1:T} = \{y_1,\ldots,y_T \}$. Since there is no closed form solution exists for extracting these information about the state variables and parameters, we consider Monte Carlo based approximation methods. 

The sequential and dynamic nature of SSMs suggests to use sequential Monte Carlo (SMC) methods, namely particle filters are widely used to learn latent state variables from the data when the model parameters are assumed to be known. When model parameters are also unknown, for simultaneous state and parameter estimation Particle  Markov  chain  Monte  Carlo  (PMCMC) \cite{andrieu2010particle} techniques have been established. 
PMCMC methods are a (non-trivial) combination of MCMC and SMC methods, where SMC algorithms are used to design efficient high dimensional proposal distributions for MCMC algorithms. The two main techniques in the PMCMC framework are Particle Metropolis Hastings (PMH) sampler and particle Gibbs (PG) samplers. We mainly focus on Particle Gibbs methods.

Particle Gibbs (PG) is a PMCMC algorithm that mimics the Gibbs sampler. In PG, samples from the joint posterior are generated by alternating between sampling the states and the parameters. 
Two major drawbacks of PG is path degeneracy and computational complexity.
When the number of states and parameters is large, the PMCMC algorithms can become computationally inefficient. We discuss various extensions of PG sampler that address one or both of the problems (path degeneracy and computational complexity). The extension of PG, particle Gibbs with ancestor sampling (PGAS), alleviates  the  problem  with  path  degeneracy and reduces the computational cost from quadratic to linear in the number of timesteps, $T$, in favorable conditions. Interacting particle Markov chain Monte Carlo (iPMCMC) \citep{rainforth2016interacting} was introduced to mitigate the path degeneracy problem, by using trade-off between exploration and exploitation that resulted in improved mixing of the Markov chains. 
Blocked Particle Gibbs (bPG) Sampler \cite{singh2017blocking} addresses the time complexity problem, by dividing the whole sequence of states into small blocks,  such that some blocks (odd or even) can be computed in parallel.

PG is an exact approximation of the Gibbs sampler and can never do better than the Gibbs sampler it approximates. To improve its performance beyond the underlying Gibbs sampler, collapsed particle Gibbs was proposed in \cite{wigren2019parameter}. In  collapsed PG, one or more parameters are marginalized over when the parameter prior is conjugate to the complete data likelihood.

Each method discussed above are implemented in Python and Rust programming languages.  Python is a general purpose programming language and is known for its simple syntax and readable code.  Rust is a systems programming language, its compile-time correctness guarantees the fast performance. We compare run-time performance of Rust and Python programs for particle Gibbs methods.

The rest of the paper is organized as follows. 
We start with an introductory background on state-space models and Monte Carlo methods in Section 2. 
Then we introduce Particle Gibbs method in Section 3. 
In Section 4, we discuss extensions and variants of PG methods.
In Section 6, we conclude and discuss future outlook.



\section{Background} \label{sec:models}
In this section, we provide a brief background on state-space models, Monte Carlo methods and we fix notations and assumptions used throughout the paper. Here we only provide a brief review of the underlying principles of MCMC and SMC methods in terms of usage of them for inference problems associated with SSMs. There is an extremely rich literature on Monte Carlo methods: see for example \citep{gelman1992inference} and \citep{gilks1995markov}.

\subsection{State Space Models}
State Space Models (SSMs) have been widely used in a variety of fields, for example, econometrics~\cite{nonejad2015particle}, ecology~\cite{parslow2013bayesian},  climatology~\citep{calafat2018coherent},  robotics~\citep{deisenroth2013survey}, and epidemiology~\citep{rasmussen2011inference}, to mention just a few.

Usually, in a SSM there is an unobserved state of interest $x_t$ that evolves through time, however, only noisy or partial observations of the state $y_t$ are available. The state process is assigned an initial density $ x_1 \sim p(x_1 | \theta)$, and evolves in time with transition density $p(x_{t} | x_{t-1})$. Given the latent states $x_t$, the observations are assumed to be independent with density $p(y_{t} | x_t, \theta)$. Here, $\theta$ is a parameter vector with prior density $p(\theta)$. 
The SSM can be expressed in probabilistic form as follows.
\begin{subequations}
\begin{align}
    x_1 & \sim p(x_1 | \theta) \\
    x_{t} | x_{t-1}, \theta  & \sim p(x_{t} | x_{t-1}, \theta)\\
    y_{t} | x_t, \theta & \sim p(y_{t} | x_t, \theta),\\
    \theta & \sim p(\theta),
\end{align}
\end{subequations}

We assume in all our experiments that the initial state is always fixed and the other model parameters $\theta = (Q, R)^T$ are fixed for some setting only.
Using the Markov Property and conditional probabilities, the joint distribution $p(x_{1:T}, \theta, y_{1:T})$ can be factorized as follows.
\begin{align}
    p(x_{1:T}, \theta, y_{1:T}) = \left( \prod_{t=1}^T p(y_{t} | x_{t}, \theta) \right) \; \left( \prod_{t=2}^T p(x_{t} | x_{t-1}, \theta) \right) \; p(x_1 | \theta) \;  p(\theta)
\end{align}

\begin{figure}[H]
    \centering
    \includegraphics[scale=.9]{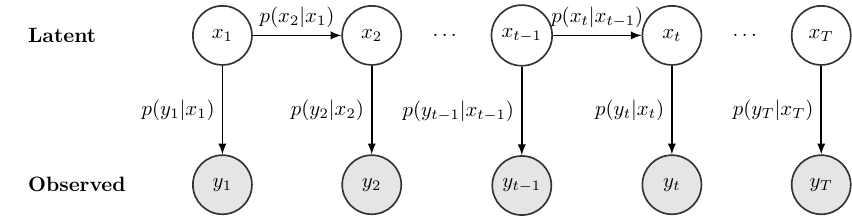}
    \caption{Graphical representation of a finite state space model. State variables $x_t$ are latent variables and measurements $y_t$ are observed variables. The probabilistic relationship between variables are shown with directed lines. Dependency on $\theta$ is omitted for simplicity}
    \label{fig:ssm_figure}
\end{figure}

The posterior distribution of the unknowns in the model can be factorized as follows:
\begin{align}
    p(x_{1:T}, \theta| y_{1:T}) =  p(x_{1:T} | \theta, y_{1:T}) \;  p(\theta|y_{1:T})
\end{align}
The term, $p(\theta|y_{1:T})$, estimation of the parameter vector $\theta$ given the observations $y_{1:T}$ is referred to as parameter inference. 
The term $ p(x_{1:T} | \theta, y_{1:T})$ is referred to as state inference problem which involves the estimation of the states $x_{1:T}$ 
given $\theta$ and $y_{1:T}$. These inference problems are  analytically intractable for most SSMs. We consider MCMC for parameter inference,  SMC for state inference and PMCMC (Particle Gibbs sampler) for simultaneous estimation of state and parameters. 

\subsection{Data Simulation from Non-linear SSM}
We simulate data from model as defined in Eq. (\ref{eq:ssm}), with the following settings.
\begin{equation}
\begin{aligned}
    f(x_t, t) & = 0.5*x_t + 25*x_t/(1+x_t^2) + 8*cos(1.2*t) \\
    g(x_t) &= x_t^2/20 \\
    \theta &= \{Q=0.1, R=1 \} \\
    x_1 &=0 \\
    T&=500 \\
\end{aligned}\label{eq:data_simulation}
\end{equation}

\begin{center}
\begin{figure}[H]
    \centering
    \includegraphics[scale=.3]{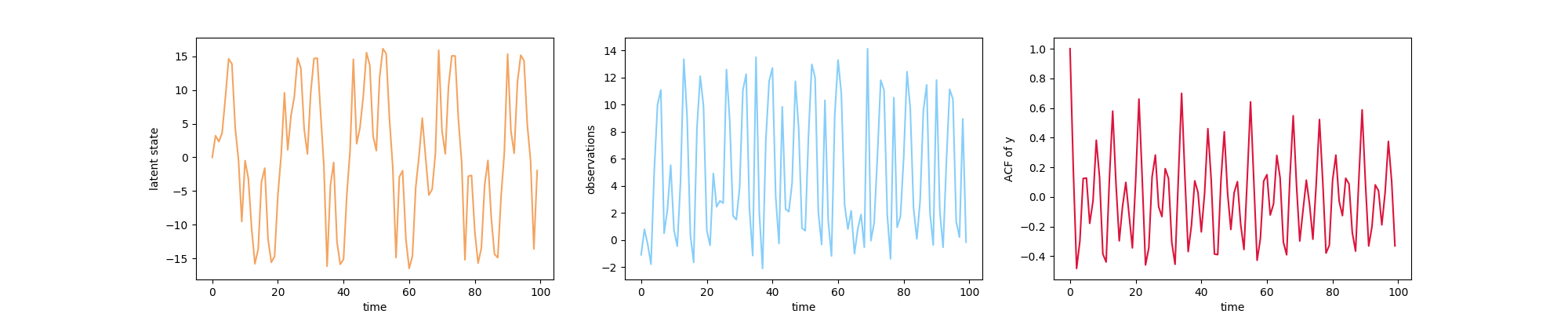}
    \caption{Simulated data from the non-linear SSM model with latent state (orange), observations (blue) and autocorrelation function (ACF) of the observations (coral).}
    \label{fig:data}
\end{figure}
\end{center}

The simulated data for first 100 time points is plotted in Figure \ref{fig:data}. We use this data through out the paper for various experiments.

\subsection{Parameter Inference using Sampling Methods}
We consider a Bayesian approach to parameter estimation, using Markov chain Monte Carlo (MCMC) methods which are based on simulating a Markov chain with the target
as its stationary distribution, $p(\theta|y_{1:T})$. Efficient and broadly used MCMC methods are:  the Metropolis-Hastings and Gibbs sampler and their variants.
Here, we consider the Gibbs sampler as proposed in~\cite{geman1993stochastic}. Gibbs sampler updates a single parameter at a time by sampling from the conditional distribution for each parameter given the current value of all the other parameters and repeatedly applying this updating process. For details on how and why Gibbs sampler work we recommend the tutorial~\cite{casella1992explaining}.

\begin{algorithm}[H]
\caption{Gibbs Sampler}\label{algo:gibbs}
\begin{algorithmic}[1]
\State \textbf{Initialize}{ set $\theta[1]$: arbitrarily}
\For{$m=2$ to number of iterations M}
\State Draw $\theta_1[m] \sim p(\theta_1[m] \mid y_{1:T}, \theta_2[m-1] \ldots \theta_k[m-1]$)
\State \ldots 
\State Draw $\theta_k[m] \sim p(\theta_k[m] \mid y_{1:T}, \theta_1[m] \ldots \theta_{k-1}[m]$)
\EndFor
\end{algorithmic}
\end{algorithm}

In a SSM, sampling from $p(\theta_i \mid y_{1:T}, \theta_1 \ldots \theta_k)$ involves the likelihood $p(y|\theta)$. Since there is no closed form expression available for the likelihood $p(y|\theta)$, one can use an estimate of the likelihood. In this tutorial, we are mainly interested in the state inference problem or the simultaneous inferences of state and parameters, which is discussed in the sections below.

\subsection{State Inference using Particle Filters}
When $\theta \in \Theta$ is known sequential Monte Carlo methods (SMC) are used for inference about  states. In particular, we consider SMC methods to approximate the sequence of posterior densities  $p(x_{1:t} | y_{1:t})$ by a set of $N$ random weighted samples called particles.
\begin{align}
   \hat{p}(x_{1:t} | y_{1:t}) & =  \sum_{i=1}^{N} w_t^i \delta_{x_{1:t}}(x_{1:t}),
\end{align}
where $w_t^i$ is a importance weight associated with particle $x_{1:t}^i$.

There are broadly two types of state inference problems in SSMs, filtering and smoothing. We mainly focus on inference problems related with marginal filtering, in which observations $y_{1:t}$  up to the current time step $t$ are used to infer the current value of the state $x_t$.
Bayesian filtering recursions are used iteratively to solve the filtering problem  for each time $t$ by using the following two steps.
\begin{align}
    p_\theta(x_{t} | y_{1:t}) & =  \frac{p_\theta(x_{t+1} | y_{1:t})}{p_\theta(y_t | y_{1:t})}\\
    y_{t} | x_t & \sim g_{\theta}(y_{t} | x_t),
\end{align}

We consider the simplest particle filter called Bootstrap Particle Filter (BPF) or standard SMC. 
At a high level SMC works as follows.
At time 1,$N$ particles ${x_{1}^i}$, for $i=1, \ldots, N$, are generated from prior $p(x_1|\theta)$ and the corresponding importance weights are computed using $\Tilde{w}_1^i = p(y_1|\theta, x_1^i)$. To generate $N$ particles approximately distributed according to the posterior $p(x_1|\theta)$ we sample $N$ times from the Importance Sampling (IS) approximation $\hat{p}(x_1|y_1)$, this is known as resampling step. At time 2 the algorithm aims to produce samples approximately distributed according to $p(x_{1:2}|\theta, y_{1:2})$ using the samples obtained at time 1. This process is then repeated for $T$ times. 
The standard particle filter is summarized in  Algorithm~\ref{algo:bpf}, where $Cat$ denotes categorical distribution. 
 We refer to \citep{doucet2009tutorial} for a gentle introduction on SMC technique.
 
\begin{algorithm}[H]
\caption{SMC}\label{algo:bpf}
\begin{algorithmic}
\State \textbf{Initialize}{}
\State Draw $x_1^i \sim p(x_{1} | \theta)$ for $i = 1 \ldots N$
\State Compute $\Tilde{w}_1^i = p(y_1|\theta, x_1^i)$ and normalize $w_1^i = \Tilde{w}_1^i/ \sum \Tilde{w}_1^i$  for $i = 1 \ldots N$
\For{$t=2$ to number of states T}
\State Sample ${a_t}^i = Cat(w_{t-1}^1, ..., w_{t-1}^N)$ for $i = 1 \ldots N$ and set $\bar{x}_{t-1} = x_{t-1}^{a_t}$
\State Draw $x_t^i \sim p(x_{t} | \bar{x}_{t-1}^i, \theta)$ for $i = 1 \ldots N$
\State Set $x_{1:t} = \{ x_{1:t-1}, x_t \} $
\State Compute $\Tilde{w}_t^i = p(y_1|\theta, x_1^i)$ and normalize $w_t^i = \Tilde{w}_t^i/ \sum \Tilde{w}_t^i$, for  $i = 1 \ldots N$
\EndFor
\end{algorithmic}
\end{algorithm}

\subsection*{Error in the latent state estimation using SMC}
\begin{figure}[H]
    \centering
    \includegraphics[scale=0.5]{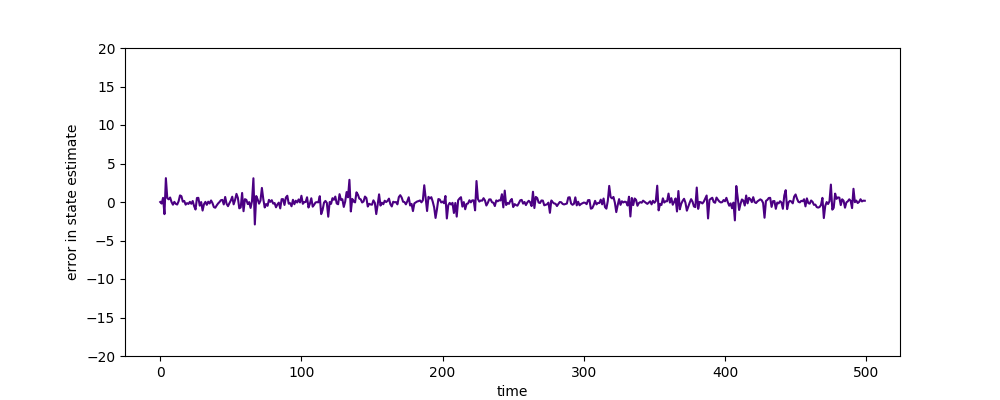}
    \caption{Error in the latent state estimate using SMC with $N=500$}
    \label{fig:bpf}
\end{figure}

 Consider the data generated from \ref{eq:data_simulation}. The difference between the true states and the estimated states using SMC with $N=500$ is plotted in Figure~\ref{fig:bpf}.
 
\subsection{State and Parameter Estimation using Particle Gibbs}
Intuitively, Gibbs sampler for simultaneous state and parameter inferences in SSMs can be thought of as alternating between updating $\theta$ and updating $x_{1:T}$:
\begin{align}
    \text{Draw } \theta[m] & \sim p(\theta \mid x_{1:T}[m-1], y_{1:T}) \\
    \text{Draw } X_{1:T}[m] & \sim p(x_{1:T}[m-1] \mid \theta[m], y_{1:T}).
\end{align}
     
However, it is hard to draw from $p( x_{1:T}[m-1] \mid \theta[m], y_{1:T})$. Therefore, we approximate $p(x_{1:T}[m-1] \mid \theta[m], y_{1:T})$ using particle filter. More specifically, we use a conditional SMC (cSMC) for which one pre-specified path is retained throughout the sampler. cSMC and PG are discussed in details in the next section.

\section{Particle Gibbs Method}
The particle Gibbs (PG) sampler was introduced in \citep{andrieu2010particle} as a way to use the approximate SMC proposals within exact MCMC algorithms (Gibbs sampler). It has been widely used for joint parameter and state inference in non-linear state-space models. 
First, we define conditional particle filter (cSMC), which is the basic building block of PG methods.

\subsection*{Conditional SMC}
Conditional SMC (cSMC) or Conditional Particle Filters (CPF) is similar to a standard SMC algorithm except that a pre-specified path, $x'_{1:t}$, is retained to all the resampling steps, whereas the remaining $N - 1$ particles are generated as usual.
For simplicity, we set the last ($N^{th}$) particle $x_t^N = x'_t$ and its ancestor index $a_t^N = N$ deterministically, where $N$ is the number of particles. Here, conditioning ensures correct stationary distribution for any $N \geq 2$. 
The cSMC algorithm returns a trajectory, indexed by $b$, where $b$ is sampled with probability proportional to the final particle weights, $b \sim Cat(\{w_T^i\}_{i=1}^N)$.
The cSMC algorithm is summarized in Algorithm \ref{algo:cpf}.

\begin{algorithm} [H]
\caption{cSMC}\label{algo:cpf}
\begin{algorithmic}
    \State \textbf{Initialize}{}
    \State Draw $x_1^i \sim p(x_{1} | \theta)$ for $i = 1 \ldots N-1$ and 
    and set $x_1^N = x'_1$
    \State Compute normalized weights $w_1^i$  for $i = 1 \ldots N$
    \For{$t=2$ to number of states T}
        \State Sample ${a^i_t}$ for $i = 1 \ldots N-1$, and set $a^N_t = N$
         and $\bar{x}_{t-1} = x_{t-1}^{a_t}$
        \State Draw $x_t^i \sim p(x_{t} |\bar{x}_{t-1}^i, \theta)$ for $i = 1 \ldots N-1$ and set $x_t^N = x'_t$
        \State Set $x_{1:t} = \{ x_{1:t-1}, x_t \} $
        \State Compute normalized weights $w_t^i$  for $i = 1 \ldots N$
    \EndFor
    \State Draw $b \sim Cat(\{w_T^i\}_{i=1}^N)$ \\
    \Return $x_{1:T}^b$
\end{algorithmic}
\end{algorithm}

The PG algorithm iteratively runs cSMC  sweeps  as  shown  in  Algorithm  \ref{algo:pg},  where  each conditional trajectory is sampled from the surviving trajectories of the previous sweep. 

\begin{algorithm} [H]
\caption{PG}\label{algo:pg}
\begin{algorithmic}[1]
\State \textbf{Initialize}{ set $x_{1:T}[1]$ and $\theta[1]$: arbitrarily}
\For{$m=2$ to number of iterations, M}
\State Draw $\theta[m] \sim p(. \mid x_{1:T}[m-1], \theta[m-1]$)
\State $X_{1:T}[m]$ = cSMC$(x_{1:T}[m-1], \theta[m], y_{1:T})$
\EndFor
\end{algorithmic}
\end{algorithm}

\subsubsection*{Simultaneous State and Parameter Inference using Particle Gibbs}
\begin{figure}[H]
    \centering
    \includegraphics[scale=0.5]{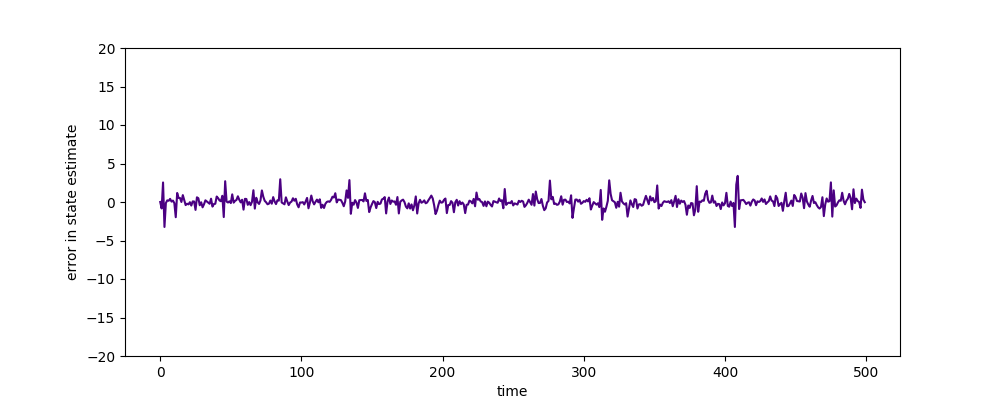}
    \caption{Error in the latent state estimate using PG with 500 particles and 50000 iterations}
    \label{fig:pg_state_x}
\end{figure}

\begin{figure}[H]
    \centering
    \includegraphics[scale=0.6]{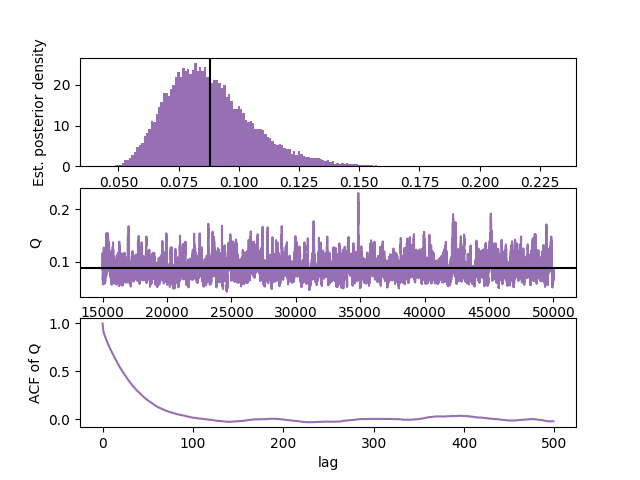}
    \caption{Posterior Q}
    \label{fig:pg_Q}
\end{figure}

\begin{figure}[H]
    \centering
    \includegraphics[scale=0.6]{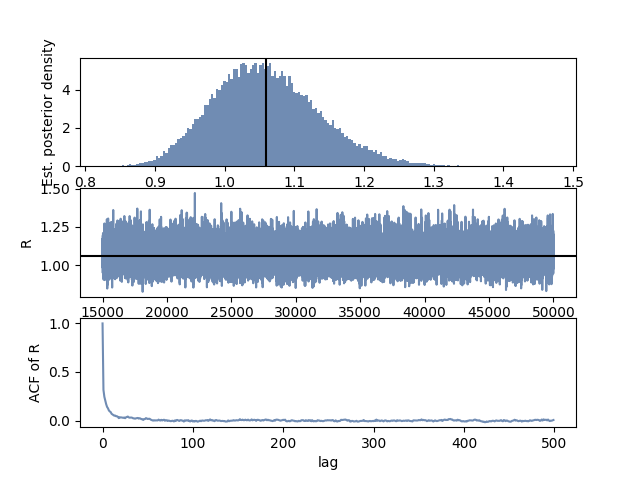}
    \caption{Posterior R}
    \label{fig:pg_R}
\end{figure}

In this experiment, we use a dataset simulated from \ref{eq:data_simulation} and the model parameters and latent states are assumed to be unknown. The error in the latent state estimate using PG with $500$ particles and $50,000$ iterations is plotted in  Figure \ref{fig:pg_state_x}. The parameter posteriors (after discarding the first one third of the samples as burn-in) are plotted in Figures \ref{fig:pg_Q} and \ref{fig:pg_R}.

\subsubsection*{Comparing Run-time Performance of Python and Rust Programs for PG}
For the previous example, the run-time performance of Python and Rust programs for PG sampler against different number of iterations (with fixed $N=500$) are given in Table 1  and are plotted in Figure \ref{fig:time_pg}. The table shows that the Rust program is 10 times faster than Python program.

\begin{minipage}{\textwidth}
  \begin{minipage}[b]{0.49\textwidth}
    \centering
    \begin{figure}[H]
        \centering
        \includegraphics[scale=0.25]{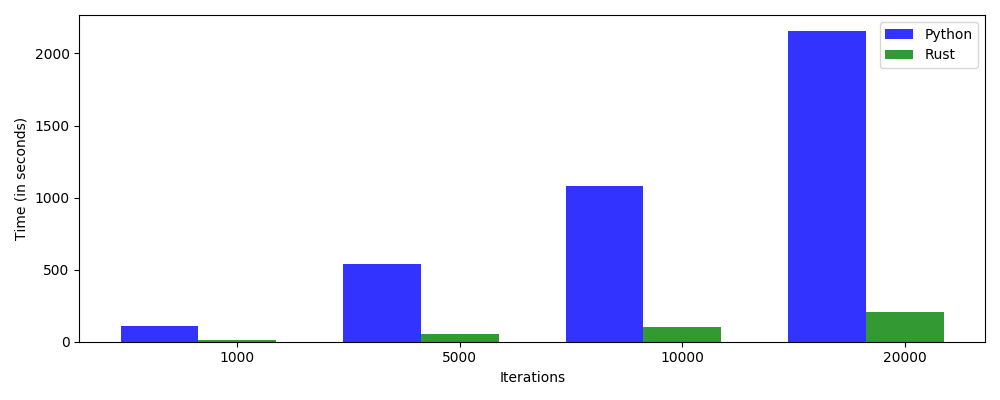}
        \caption{Visualizing run-time performance of Python and Rust program for PG against different number of iterations. }
        \label{fig:time_pg}
    \end{figure}
  \end{minipage}
  \hfill
  \begin{minipage}[b]{0.4\textwidth}
    \centering
    \begin{tabular}{ccc}\hline
      \# Iters &Python & Rust \\ \hline
       1000 & 109 & 10 \\
        5000 & 538 & 52 \\ 
        10000 & 1079 & 102 \\ 
        20000 & 2159  & 204 \\ 
        \hline
      \end{tabular}
      \label{tab:time_pg}
      \captionof{table}{Comparison of time (in seconds) for Python and Rust program for PG.}
    \end{minipage}
  \end{minipage}

\section{Extensions and Variants of Particle Gibbs Methods}

In this section, we discuss various extensions and variants of particle Gibbs method.

\subsection{Particle Gibbs with Ancestor Sampling}
PG algorithm has been proven to be uniformly ergodic under standard assumptions, however, the mixing of the PG sampler can be poor, especially when there is severe degeneracy in the underlying SMC. It has been shown that the number of particles $N$ must increase linearly with $T$ for the sampler to mix properly for large $T$, which results in an overall quadratic computational complexity with $T$. To address this problem PGAS was introduces in \cite{lindsten2014particle}. In PGAS, the ancestor for the reference trajectory in each time step is sampled, according to ancestor weights, instead of setting it deterministically, which significantly improves the mixing of the sampler for small $N$, even when $T$ is large. 

Mainly, ancestor resampling within cSMC was introduced to mitigates path degeneracy and that helps in movement around the conditioned path. Instead of setting $a_t^N = N$, a new value is sampled from $\{ 1 \ldots T\}$. The idea is to connect the partial reference trajectory $x'_{t:T}$ to one of the particles $x^i_{1:t-1}$. It is done in the following two steps:
\begin{align}
    \textbf{Compute weights: } & \Tilde{w}_{t-1|T} \propto w^i_{t-1} \;  p(x'_t|x^i_{t-1}) \\
     \textbf{Sample : } &  P(a_t^N = i) \propto \Tilde{w}_{t-1|T}
\end{align}

The cSMC-AS algorithm is summarized as follows.
\begin{algorithm} [H]
\caption{cSMC-AS}\label{algo:cSMC_as}
\begin{algorithmic}
    \State \textbf{Initialize}{}
    \State Draw $x_1^i \sim p(x_{1}|\theta)$ for $i = 1 \ldots N-1$ and 
    and set $x_1^N = x'_1$
    \State Compute normalized weights $w_1^i$  for $i = 1 \ldots N$
    \For{$t=2$ to number of states T}
        \State Sample ${a^i_t}$ for $i = 1 \ldots N-1$, 
        \State \textbf{Compute weights: }  $\Tilde{w}_{t-1|T} \propto w^i_{t-1}  p(x'_t|x^i_{t-1}) $
        \State \textbf{Sample : } $  P(a_t^N = i) \propto \Tilde{w}_{t-1|T}$
        \State $\bar{x}_{t-1} = x_{t-1}^{a_t}$
        \State Draw $x_t^i \sim p(x_{t} | \bar{x}_{t-1}, \theta)$ for $i = 1 \ldots N-1$ and set $x_t^N = x'_t$
        \State Set $x_{1:t} = \{ x_{1:t-1}, x_t \} $
        \State Compute normalized weights $w_t^i$  for $i = 1 \ldots N$
    \EndFor
    \State Draw $b \sim Cat(\{w_T^i\}_{i=1}^N)$ \\
    \Return $x_{1:T}^b$
\end{algorithmic}
\end{algorithm}

The PGAS algorithm is the same as PG except the step cSMC in PG is replaced with cSMC-AS in PGAS.
\begin{algorithm}[H]
\caption{PGAS}\label{algo:pgas}
\begin{algorithmic}[1]
\State \textbf{Initialize}{ set $x_{1:T}[1]$ and $\theta[1]$: arbitrarily}
\For{$m=2$ to number of iterations, M}
    \State Draw $\theta[m] \sim p(. \mid x_{1:T}[m-1], \theta[m-1]$)
    \State $X_{1:T}[m]$ = cSMC-AS$(x_{1:T}[m-1],\ldots)$
\EndFor
\end{algorithmic}
\end{algorithm}

\subsubsection*{Mixing of PG and PGAS}
\begin{figure}[ht]
\begin{subfigure}{.5\textwidth}
  \centering
  \includegraphics[scale=0.3]{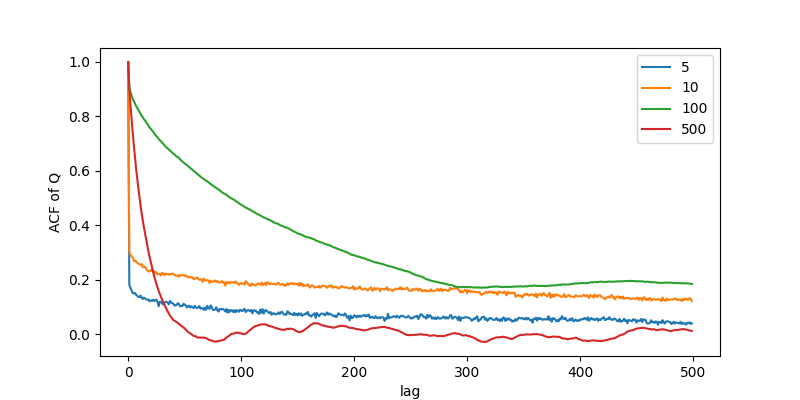}  
  
\end{subfigure}
\begin{subfigure}{.5\textwidth}
  \centering
  \includegraphics[scale=0.3]{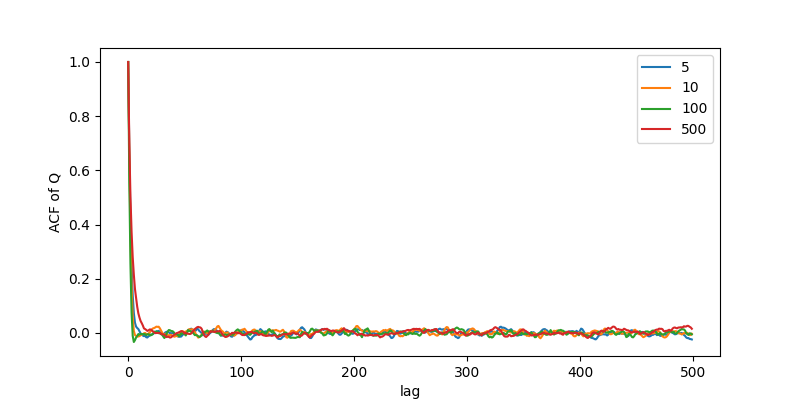}  
  
\end{subfigure}
\caption{ACFs of the parameter $Q$ for PG (left column) and for PGAS (right column) for a dataset generated from \ref{eq:data_simulation} with $T=500$.  The results are reported against different number of particles $N$.}
\label{fig:acf}
\end{figure}

 To illustrate that ancestor resampling can considerably improve the mixing of PG, we plot AutoCorrelation Functions (ACF) of the noise parameter $Q$. We consider the dataset generated from \ref{eq:data_simulation}, for $T=500$, and we assume that the model parameters and states are unknown. The PG and PGAS samplers are simulated for $50000$ iterations, and the first one third of the samples are discarded as burn-in. The ACFs of $Q$ for PG and PGAS against different values of $N= (10, 50, 100, 500)$ are plotted in Figure \ref{fig:acf}, which show that PG sampler performs poorly for smaller $N$ ($N=5, 10$), and large $N(>100)$ is required to obtain good mixing. However, PGAS is much more robust, even for small $N$ it shows good mixing rates.

\subsection{Interacting particle Markov chain Monte Carlo.}
As mentioned in previous section, a major drawback of PG is path degeneracy in cSMC step, where conditioning on an existing trajectory implies that whenever resampling of the trajectories results in a common ancestor, this ancestor must correspond to the reference trajectory.   This results in high correlation between the samples, and poor mixing of the Markov chain. 
Interacting particle Markov chain Monte Carlo (iPMCMC) \citep{rainforth2016interacting} was introduce to mitigate this problem by, time to time switching between a cSMC particle system with a completely independent SMC one, which results in improved mixing.

In iPMCMC a pool of conditional SMC samplers and standard SMC samplers are run as parallel processes, where each process is referred to as node. Assume that there are $R$ separate nodes, $P$ of them run cSMC and $R-P$ run SMC, and they can interact by exchanging only very minimal information at each iteration to draw new MCMC samples.  The cSMC nodes are given an identifier $c_j \in \{ 1, \ldots, R \}$, where $j \in \{ 1, \ldots, P \}$. Let $x^i_r = x^i_{1:T, r} $ be the internal particle trajectories of node $r \in \{ 1, \ldots, R \}$. At each iteration $m$, the nodes $c_{1:P}$ run cSMC with the previous MCMC samples $x'_j[r-1]$ as the reference particle. The remaining $R-P$ nodes run standard SMC.  Each node $r$ returns an estimate of the marginal likelihood for the internal particle system defined as
\begin{align}
    \hat{Z}_r  = \prod_{t=1}^{T} \sum_{i=1}^{N} w^i_{t,r}
\end{align}

The new conditional nodes are then set by sampling new indices $c_j$ as follows.
\begin{align}
    p(c_i =r| c_{1:P\setminus j}) & = \hat{\zeta}^j_r \label{eq:sample_c_j} \\
    \hat{\zeta}^j_r & = \frac{ \hat{Z}_r \mathcal{I}(r \notin c_{1:P\setminus j})}{\sum_q \hat{Z}_q \mathcal{I}(q \notin c_{1:P\setminus j})}
\end{align}
Thus one loop through the conditional SMC node indices is required to resample them from the union of the current node index and the unconditional SMC node indices, in proportion to their marginal likelihood estimates. This is the key step that may switch the nodes from which the reference particles will be drawn.

\begin{algorithm}[H]
\caption{iPMCMC}\label{algo:iPMCMC}
\begin{algorithmic}[1]
\State \textbf{Input}: number of nodes: R, conditional nodes: P, and MCMC steps: M
\State \textbf{Initialize}{ set $x'_{1:P}[1]$ 
}
\For{$m=2$ to number of iterations, M}
\State Workers $c_{1:P}$ run cSMC using $x'_{1:P}[m-1]$ as reference particles
\State Workers $1:R \setminus c_{1:P}$ run SMC 
    \For{$j=1$ to P}
    \State Simulating $c_j$ according to Eq. \ref{eq:sample_c_j}
    \State $x'_{j}[r] = x_{c_j} $
    \EndFor
\EndFor
\end{algorithmic}
\end{algorithm}

The run time performance of Rust and Python programs for iterated PG sampler is compared in Table \ref{tab:i_pg} against different number of iterations and fixed number of particles $N=500, R=16$, and $P=8$. Each program used the same data set generated from \ref{eq:data_simulation}. The table shows that the Rust program is more than 8 times faster than the Python program.

\begin{table}[H]
    \centering
     \begin{tabular}{ccc}\hline
      \# Iters &Python & Rust \\ \hline
       1000 & 486 & 56    \\
        5000 & 2436 & 282 \\ 
        10000 & 4952 & 564 \\ 
        20000 & 10128  & 1145 \\ 
        \hline
      \end{tabular}
      \caption{Comparison of time (in seconds) between Python and Rust programs for iterated PG sampler.}
    \label{tab:i_pg}
\end{table}



\subsection{Blocked Particle Gibbs Sampler}
The uniform ergodicity of the Markov kernel used in PG was proven in \cite{chopin2015particle}, and it was shown that the mixing rate does not decay if the number of particles grows at least linearly with the number of latent states. However, the computation complexity of a PG  sampler is quadratic in the number of latent states, which can be a limiting factor for its use in long observation sequences. Blocked Particle Gibbs (bPG) Sampler was introduced in \cite{singh2017blocking} to address this problem, and it was shown that using blocking strategies, a sampler can achieve a stable mixing rate for a linear cost per iteration. The main idea in Blocked PG is to divide the whole sequence of states into small (overlapping) blocks,  such that odd and even blocks can be computed in parallel.

\begin{figure}[H]
    \centering
    \includegraphics[scale=0.5]{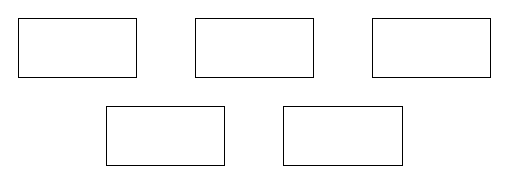}
    \caption{Blocked Particle Gibbs Strategy, where the blocks in first row are odd blocks and the second row contains even blocks, and consecutive odd and even blocks are overlapping.}
    \label{fig:bPG_strategy}
\end{figure}

Let $I = \{ 1, \ldots, T \}$ be the index set of the sequence of latent variables $X_1, \ldots, X_T$.
In blocked PG, the sequence $X_1, \ldots, X_T$ is divided into blocks, where consecutive blocks may overlap but nonconsecutive blocks do not overlap and are separated, as illustrated in Figure \ref{fig:bPG_strategy}. The block size $L$ and overlap $p$ are chosen such that the ideal sampler is stable, and the number of particles is large enough to obtain a stable PG. Note that blocked PG depends only on size of $L$ not on $T$. 

Let $\mathcal{J}={J_1, \ldots, J_r}$ be a cover of $\{1, \ldots, T\}$ and let $\mathcal{P}=P_{J_1}, \ldots, P_{J_r}$ be the Gibbs kernel for one complete sweep from left to right. The parallel Gibbs kernel are defined as follows. For simplicity we can assume that the number of blocks are even (it is easy to construct similar arguments for the odd number of blocks as well).
\begin{align*}
    \mathcal{P}_{odd} &=P_{J_1}, P_{J_3}, \ldots, P_{J_{r-1}} \\
    \mathcal{P}_{even} &=P_{J_2}, P_{J_4}, \ldots, P_{J_{r}} 
\end{align*}

In the first iteration, we sweep through the blocks from left to right, and let $\mathcal{P}$ be the kernel corresponding to one complete sweep. Then at each iteration we update all the odd-numbered blocks first and then all the even-numbered blocks. It is called parallel blocked Gibbs sampling. The kernel for an internal block $J = \{s, \ldots, u \}$, called blocked conditional SMC sampler, is defined in the following.

\subsubsection*{Blocked cSMC}
The blocked SMC approximates the sequence of target distributions 
\[ p(x_s,\ldots, x_t|x_{s-1}, y_s, \ldots, y_t)  \] for $t = s, \ldots, u$
using conditional SMC. For initialization, the distribution $p(x_{s} | x_{s-1})$ is used. The for loop of blockedSMC algorithm is similar to cSMC. After the loop, to take into account that the target distribution is  $p(x_s,\ldots, x_u|x_{s-1}, x_{u+1}, y_s, \ldots, y_t)$,  the conditioning on the fixed boundary state $ x_{u+1}$ is applied , which contributes via the term $p(x_{u+1}|X^i_u)$ to the final weight $w_u^i$. 

\begin{algorithm} [H] 
\caption{blockedSMC}\label{algo:blockedSMC}
\begin{algorithmic}
    \State \textbf{Initialize}{}
    \State Draw $x_s^i \sim p(x_{s} | x_{s-1})$ for $i = 1 \ldots N-1$ and 
    and set $x_s^N = x'_1$
    \State Compute normalized weights $w_1^i$  for $i = 1 \ldots N$
    \For{$t=s+1$ to u}
        \State Sample ${a^i_t}$ for $i = 1 \ldots N-1$, and set $a^N_t = N$
         and $\bar{x}_{t-1} = x_{t-1}^{a_t}$
        \State Draw $x_t^i \sim p(x_{t} | \bar{x}_{t-1}, \theta)$ for $i = 1 \ldots N-1$ and set $x_t^N = x'_t$
        \State Set $x_{1:t} = \{ x_{1:t-1}, x_t \} $
        \State Compute normalized weights $w_t^i$  for $i = 1 \ldots N$
    \EndFor
    Set $w_u^i = w_u^i * p(x_{u+1}|X_u)$ for $i = 1 \ldots N$.
    \State Draw $b \sim Cat(\{w_u^i\}_{i=1}^N)$ \\
    \Return $x_{s:u}^b$
\end{algorithmic}
\end{algorithm}

Note that for the first block we have deterministic initial condition as before and for the last block we do not have to adjust for the (overlapping) consecutive next block. The blocked PG is summarized in the following algorithm.

\begin{algorithm}[H]
\caption{blocked PG}\label{algo:bPG}
\begin{algorithmic}[1]
\State \textbf{Input}: size of block: L, overlap size: p, and MCMC steps: M
\State Iteration 1: Initialization: set $x_{1:T}[1]$ by calling SMC
\State Compute start and end index for each block
\For{$m=2$ to number of iterations, M}
\State Compute an initial state, a boundary state and a reference trajectory $x_{s:u}$ for each block using $x_{1:T}[m-1]$
\State Run blockedSMC for each odd block in parallel
\State Run blockedSMC for each even block in parallel
\State Combine results from all the odd and even sweeps into $x_{1:T}[m]$
\EndFor
\end{algorithmic}
\end{algorithm}

\subsubsection*{Comparing the True States and the Estimated Latent States}
\begin{figure}[H]
    \centering
    \includegraphics[scale=0.5]{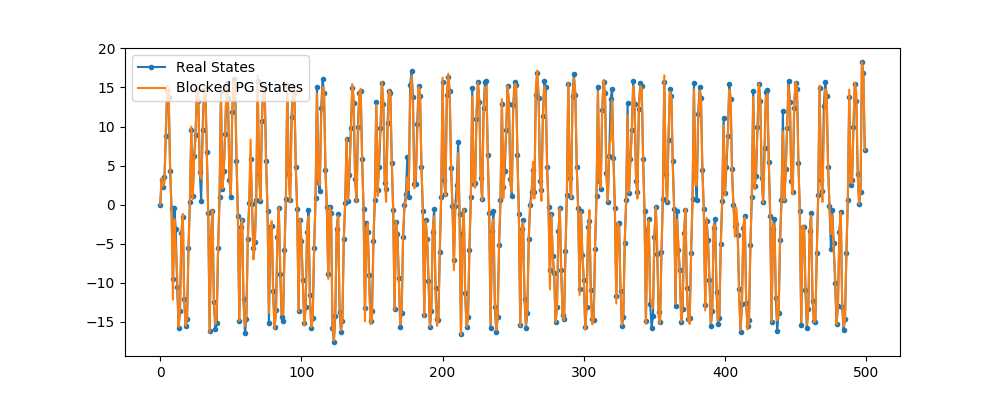}
    \caption{Comparison between the true states and the estimated states using blocked PG with $N=500$ particles and $10,000$ iterations}
    \label{fig:bPG_state}
\end{figure}

In this experiment, we compare between the true states and the estimated latent states using blocked PG for the simulated data from \ref{eq:data_simulation}, as shown in Figure \ref{fig:bPG_state}. We simulated blocked PG for $10,000$ iterations with the number of particles $N=500$, block size=30, 1 overlapping particle and all the blocks were run in parallel.  The estimated states seem to be a close estimate of the true states from Figure \ref{fig:bPG_state},.

\subsubsection*{Run-time Performance Comparison}

For the previous example, the run time performance of the Rust and Python programs for blocked PG sampler is compared in Table \ref{tab:blocked_pg} against different number of iterations and fixed number of particles $N=500$. For each sampler we used block size=30  and 1 overlapping particle and all the blocks were run in parallel. The table shows that the Rust program is almost 8 times faster than the Python program.

\begin{table}[H]
    \centering
     \begin{tabular}{ccc}\hline
      \# Iters &Python & Rust \\ \hline
       1000 & 81 & 1 \\
        5000 & 402 & 6 \\ 
        10000 & 809 & 12 \\ 
        20000 & 1625  & 25 \\ 
        \hline
      \end{tabular}
      \captionof{table}{Comparison of time (in seconds) for Python and Rust programs for blocked PG.}
    \label{tab:blocked_pg}
\end{table}

\subsection{Collapsed Particle Gibbs}
 Usually, independent samples from the target distribution are desired. When there is strong correlation between the variables the standard Gibbs 
sampler can generate correlated samples. In Gibbs sampler, when we integrate out (marginalizes over) one or more variables when sampling for some other variable, it is known as collapsed Gibbs sampler \cite{liu1994collapsed}.

Here we focus on marginalized state update, integrating out the model parameters. In particle Gibbs sampler, there is a dependence between the states $x_{1:T}$ and the model parameters $\theta$ which leads to correlated samples. 
By marginalizing out the parameters from the state update, the amount of auto correlation between samples can be reduced.

In the following we define marginalized SMC followed by the marginalized Particle Gibbs (collapsed particle Gibbs) and its application in the non linear state space models. For the detail we refer to \cite{wigren2019parameter}. 

\subsection*{marginalized SMC}

Marginalized conditional SMC (mcSMC) is similar to cSMC algorithm except that we integrate out the model parameters. We assume that there is a conjugacy relationship between the prior distribution $p(\theta)$ and the complete data likelihoods $p(x_{1:t},y_{1:t}|\theta)$, for $t= 1,\ldots,T$. The use of a restricted exponential family was proposed, where the log-partition function is assumed to be separable into two parts, one consisting of parameter-dependent part and the other having state-dependent part. The complete data likelihood under the restricted exponential family can be given by the following.
\begin{align*}
    p(x_t, y_t | x_{t-1}, \theta) = h_t exp\left( \theta^T s_t - A^T(\theta) r_t\right)
\end{align*}
where $A(\theta)$ is the restricted log-partition function and $r(x)$ is some function which only depends on $x$. A conjugate prior for this likelihood is
\begin{align*}
    p(\theta| \chi_0, \nu_0) = g(\chi_0, \nu_0) exp\left( \theta^T \chi_0 - A^T(\theta) \nu_0\right)
\end{align*}

The parameter posterior is given by:
\begin{align*}
    p(\theta| \chi_0, \nu_0) = \pi(\chi_{t-1}, \nu_{t-1}) 
\end{align*}
where the hyper-parameters are iteratively updated according to
\begin{align} \label{eq:hyper_param_update}
    \chi_t = \chi_0 + \sum_{k=1}^{t}s_k = \chi_{t-1} + s_t \\
    \nu_t = \nu_0 + \sum_{k=1}^{t}r_k = \nu_{t-1} + r_t
\end{align} 
With the above joint likelihood and conjugate prior, the expression for the marginal of the joint distribution of states and observations, at time $t$ can be derived in the closed form.
\begin{align*}
        p(x_t, y_t | x_{1:t-1}, y_{1:t-1}) & = \int     p(x_t, y_t | x_{t-1}, \theta) p(\theta| \chi_0 \nu_0) \\
         &=h_t \frac{g(\chi_{t-1}, \nu_{t-1})}{g(\chi_{t}, \nu_{t})}
\end{align*}
In order to compute the weights for the mcSMC under the restricted exponential family assumption, we only need to keep track of and update the hyper parameters according to Eq. \ref{eq:hyper_param_update}. The mcSMC method is summarized in Algorithm \ref{algo:mcSMC}.

\begin{algorithm} [H]
\caption{mcSMC}\label{algo:mcSMC}
\begin{algorithmic}
\State \textbf{Initialize}{}
\State Draw $x_1^i \sim p(x_{1} | \theta)$ for $i = 1 \ldots N-1$ and 
and set $x_1^N = x'_1$
\State Compute normalized weights $w_1^i$  for $i = 1 \ldots N$
\For{$t=2$ to number of states T}
\State Update hyperparameters $\chi_{t}^i, \nu_{t}^i$ for $i = 1 \ldots N$ 
\State Sample ${a^i_t}$ for $i = 1 \ldots N-1$, and set $a^N_t = N$
 and $\bar{x}_{t-1} = x_{t-1}^{a_t}$
\State Draw $x_t^i \sim p(x_{t} | \bar{x}_{t-1})$ for $i = 1 \ldots N-1$ and set $x_t^N = x'_t$
\State Set $x_{1:t} = \{ x_{1:t-1}, x_t \} $
\State Compute normalized weights $w_t^i$  for $i = 1 \ldots N$
\EndFor
\State Draw $b \sim Cat(\{w_T^i\}_{i=1}^N)$ \\
\Return $x_{1:T}^b$
\end{algorithmic}
\end{algorithm}

The collapsed PG algorithm iteratively runs mcSMC  sweeps  as  shown  in  Algorithm  \ref{algo:collapsed_pg},  where  each conditional trajectory is sampled from the surviving trajectories of the previous sweep.

\begin{algorithm} [H]
\caption{Collapsed PG}\label{algo:collapsed_pg}
\begin{algorithmic}[1]
\State \textbf{Initialize}{ set $x_{1:T}[1]$ and $\theta[1]$: arbitrarily}
\For{$m=2$ to number of iterations, M}
\State Draw $\theta[m] \sim p(. \mid x_{1:T}[m-1], \theta[m-1]$)
\State $X_{1:T}[m]$ = mcSMC$(x_{1:T}[m-1],\ldots)$
\EndFor
\end{algorithmic}
\end{algorithm}

\subsubsection*{Comparing Mixing Rate of PG and Collapsed PG}
\begin{figure}[ht]
\begin{subfigure}{.5\textwidth}
  \centering
  \includegraphics[scale=0.3]{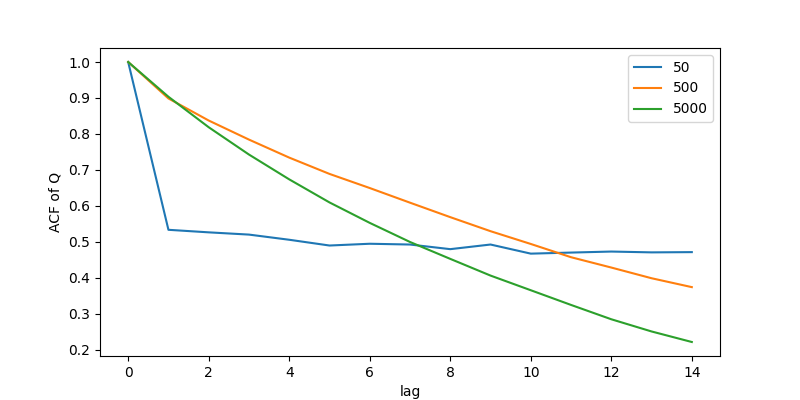}  
  
\end{subfigure}
\begin{subfigure}{.5\textwidth}
  \centering
  \includegraphics[scale=0.3]{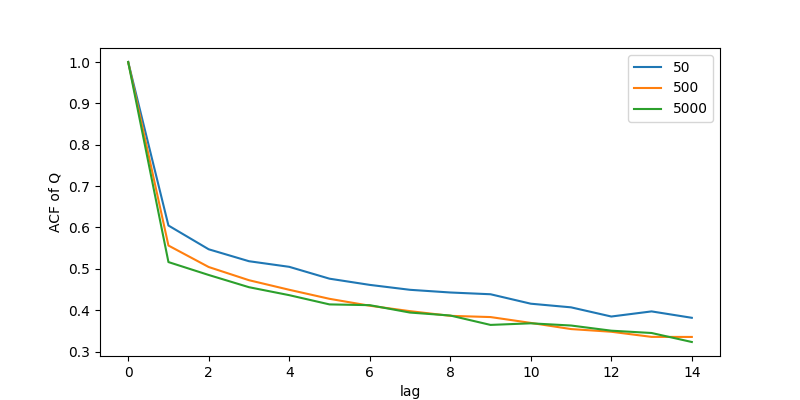}  
  
\end{subfigure}
\caption{ACFs of the parameter $Q$ for PG (left column) and for collapsed PG (right column) for a dataset simulated from \ref{eq:data_simulation}.  The results are reported against different number of particles N.}
\label{fig:collapsed_acf}
\end{figure}

To compare the mixing rate, we simulated PG sampler and collapsed PG sampler for 10,000 iteration with varying number of particles, for the dataset generated from \ref{eq:data_simulation}. After discarding the first one third samples, the first 15 lags of ACFs are computed and are plotted as shown in Figure \ref{fig:collapsed_acf}. The figure shows that the mixing rate of collapsed PG is much stable than the mixing rate of PG, even for small number of particles.

\section{Conclusion and Future Research} \label{sec:summary}
We discussed particle Gibbs Sampler and its variants and extensions such as Particle Gibbs with ancestor sampling, interacting particle MCMC, blocked PG and collapsed PG, for state and parameter inferences in non-linear SSMs. 
We illustrated all the methods with simulated datasets. Probably our implementations of all the methods discussed in Python and Rust programming language would make it easy to understand. We compared run time performance of the Python and Rust programs, the results show that the rust programs are 8 to 10 times faster than the corresponding Python programs.


The nature of PGAS is off-line in the sense given a new observation the algorithm has to be executed from scratch. The simultaneous estimation of parameters and states with an on-line approach will be more useful in dynamical system identification etc.

\section*{Code}
The source code of our implementation is available at
\url{https://github.com/niharikag/PGSampler}.
\section*{Acknowledgments}
 The computations were performed on resources provided by SNIC through Tetralith under project SNIC 2020/5-278. 
 
\bibliographystyle{plain}
\bibliography{PGAS}

\end{document}